\documentclass[journal,twoside,web]{ieeecolor}
\usepackage{generic}
\usepackage{cite}

\usepackage{etoolbox}
\makeatletter
\patchcmd{\thebibliography}{\itemsep 0pt plus pt}{\itemsep 0pt plus .3pt}{}{}
\makeatother

\usepackage{amsmath,amssymb,amsfonts}
\usepackage{algorithmic}
\usepackage{graphicx}
\usepackage{algorithm,algorithmic}
\usepackage{hyperref}
\hypersetup{
    colorlinks=true,
    linkcolor=blue,
    citecolor=black,
    urlcolor=black
}
\usepackage{textcomp}
\def\BibTeX{{\rm B\kern-.05em{\sc i\kern-.025em b}\kern-.08em
    T\kern-.1667em\lower.7ex\hbox{E}\kern-.125emX}}

\makeatletter
\def\ps@titlepagestyle{%
  \def\@oddfoot{}\def\@evenfoot{}%
  \def\@oddhead{\hfil\textsf{\scriptsize\thepage}}%
  \def\@evenhead{\textsf{\scriptsize\thepage}\hfil}%
}
\def\@oddhead{\hfil\textsf{\scriptsize\thepage}}%
\def\@evenhead{\textsf{\scriptsize\thepage}\hfil}%
\def\@oddfoot{}\def\@evenfoot{}%
\makeatother

\usepackage{booktabs}
\usepackage{tabularx}
\usepackage{multirow}

\def\refname{References}

\newcommand{\Rbb}{\mathbb{R}}
\newcommand{\Mset}{\mathcal{M}}

\begin{document}
\title{NeuroWorld: A Latent Brain World Model for Stimulus-Conditioned Human Brain Dynamics}
\author{Zijian Dong, Jianxiong Zhou, Kwun Kei Ng, Jan Paolo Macapinlac Balagtas, Zhizhou Li, Zijiao Chen, Juan Helen Zhou
\thanks{Z. Dong, J. Zhou, K. K. Ng, J. P. M. Balagtas, Z. Li, and J. H. Zhou are with the Centre for Sleep and Cognition \& Centre for Translational Magnetic Resonance Research, Yong Loo Lin School of Medicine, National University of Singapore, and with the Healthy Longevity \& Human Potential Translational Research Programme and Department of Medicine, Yong Loo Lin School of Medicine, National University of Singapore, Singapore. (email: zijian\_dong@nus.edu.sg; jianxiong.zhou@nus.edu.sg; eric.nkk@nus.edu.sg; janpaolo@nus.edu.sg; zhizhou.li@u.nus.edu; helen.zhou@nus.edu.sg)}
\thanks{J. H. Zhou is also with the Department of Electrical and Computer Engineering, National University of Singapore, Singapore.}
\thanks{Z. Chen is with the Department of Psychology, Stanford University, Stanford, CA, USA. (email: zijiao@stanford.edu)}
\thanks{Z. Dong and J. Zhou contributed equally, and are listed in random order.}
}

\maketitle

\begin{abstract}
Forecasting human brain activity during naturalistic experience requires modeling the causal evolution of endogenous neural states under continuous sensory drive. Yet existing brain encoding models largely frame this problem as stimulus-to-response regression and lack strict temporal constraints, allowing information from future stimuli to leak into current predictions. We introduce NeuroWorld, which, to our knowledge, is the first brain world model to cast naturalistic brain functional dynamics prediction as stimulus-conditioned evolution in a learned latent brain-state space. Its two-stage design separates endogenous brain states (measured via fMRI) from exogenous multimodal stimuli. Latent Dynamics Learning (LDL) jointly learns a transition-sufficient representation and causal dynamics through next-latent prediction, without reconstructing the observed fMRI signal. Latent Rollout Decoding (LRD) then freezes LDL components, rolls latent states forward autoregressively from an observed fMRI prefix, and decodes them into subject-specific whole-brain responses. We evaluate NeuroWorld on three naturalistic movie-fMRI benchmarks spanning 30 participants, including our newly collected Singapore Multimodal Imaging \& Naturalistic Dataset (SG-MIND) benchmark, which comprises 20 participants, 8,519 paired stimulus--response clips, and 140.7 person-hours of viewing. Under strictly causal stimulus access, NeuroWorld achieves state-of-the-art multi-step rollout performance on all three benchmarks, with greater robustness to long-horizon autoregressive drift, thereby supporting reliable simulation of extended brain-state trajectories. Extensive interpretability analyses further characterize the functional organization of the learned dynamics. Together, our study establishes latent-space world modeling as a principled framework for causal forecasting of human brain activity during continuous naturalistic experience. The code will be made publicly available upon acceptance.

\end{abstract}


\section{Introduction}
\label{sec:introduction}
\IEEEPARstart{L}{earning} representations of human brain dynamics, which is a central goal of computational neuroscience, requires more than mapping isolated stimuli to isolated neural responses. Under naturalistic experience, the brain behaves as a temporally extended dynamical system: a continuous stream of visual, auditory, and linguistic input drives a latent neural state, which in turn shapes how future brain activity unfolds. Naturalistic movie-watching functional MRI (fMRI) is therefore a natural setting for world modeling. The stimulus is externally observed, multimodal, and time-ordered; the fMRI signal traces a whole-brain trajectory; and the core computational problem becomes learning a state transition that forecasts future brain activity under ongoing sensory drive.

\begin{figure}[!t]
\centerline{\includegraphics[width=\columnwidth]{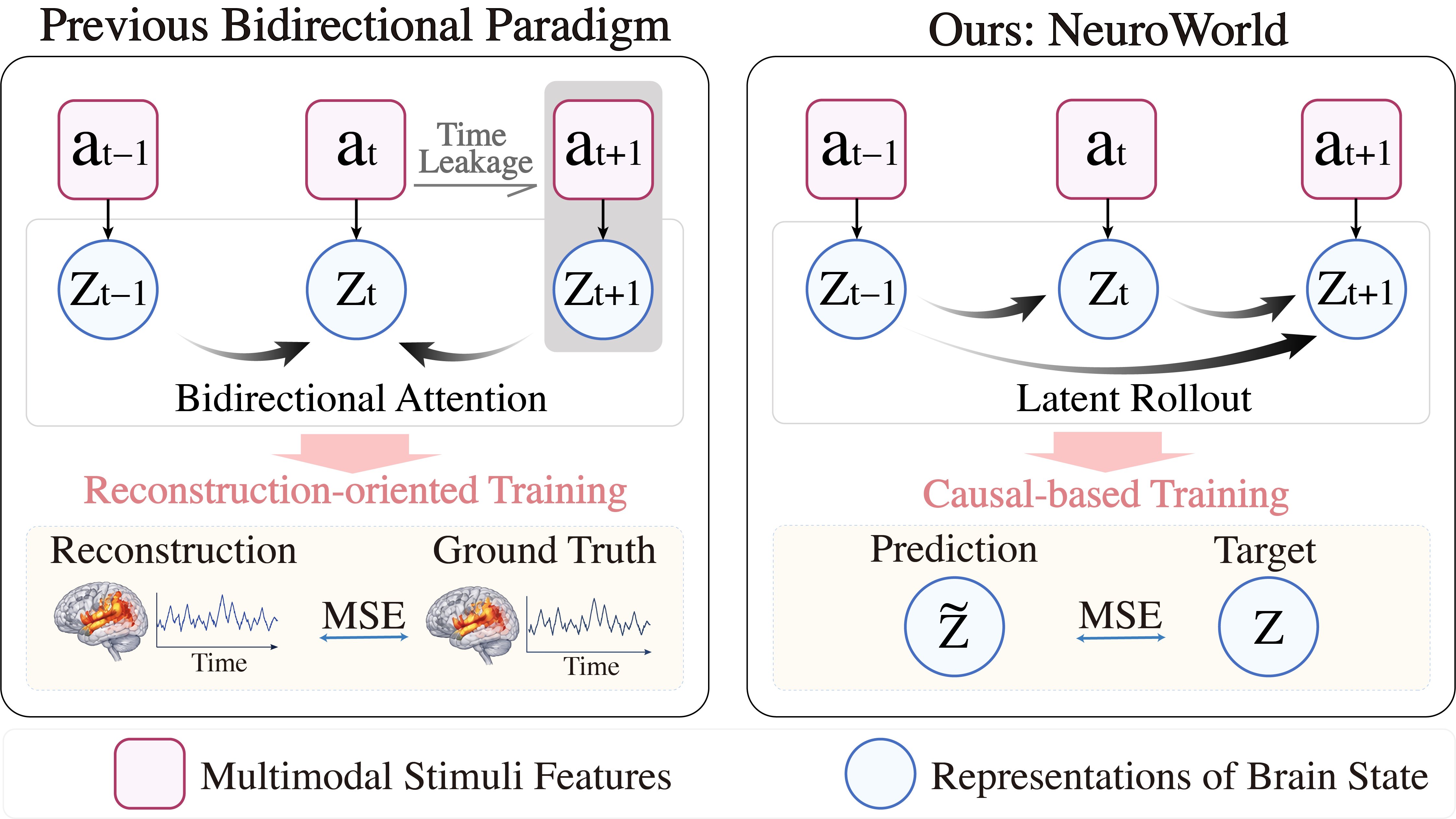}}
\caption{\textbf{Conceptual comparison between conventional bidirectional paradigms and NeuroWorld.} \textbf{Left:} Bidirectional attention permits future-stimulus leakage and learns brain representations with a reconstruction-oriented objective. \textbf{Right:} NeuroWorld enforces causal stimulus access and learns transition-sufficient latent brain states through next-latent prediction for autoregressive rollout. $a_t$ - stimulus token; $Z_t$ - latent brain state.}
\label{fig1}
\end{figure}

\begin{figure*}[!t]
\centerline{\includegraphics[width=\textwidth]{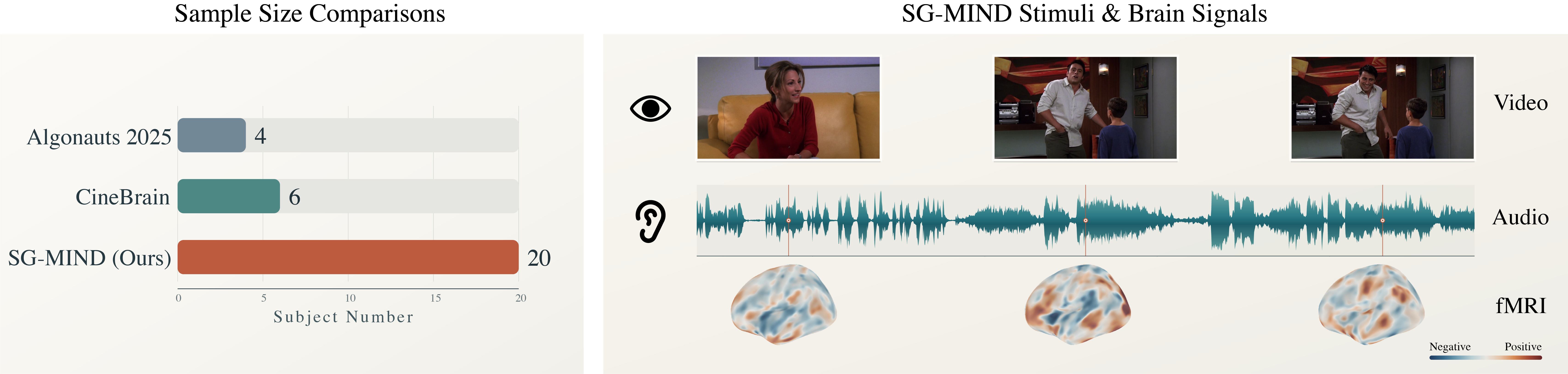}}
\caption{\textbf{Overview of the SG-MIND} \textbf{Left:} SG-MIND expands the cohort size to 20 participants, compared with four in Algonauts 2025 and six in CineBrain. \textbf{Right:} Sampled temporally aligned video, audio, and cortical fMRI signals from SG-MIND.}
\label{fig2}
\end{figure*}

This perspective reframes naturalistic brain encoding. Rather than treating encoding as a retrospective regression from stimulus onto the blood-oxygen-level-dependent (BOLD) signal in fMRI \cite{d2025tribe,d2026foundation,gokce2026mirage}, a brain world model should learn a latent brain state together with its causal evolution. This evolution must be causal in time. In other words, to predict brain activity at a given moment, the model may draw on past brain states and the movie stimulus up to that point, but not on future stimulus the brain has not yet seen. Given such a prefix, the model rolls the latent state forward and reads out the resulting trajectory (Figure~\ref{fig1}). The constraint mirrors the phenomenon itself: cortical activity at any instant is shaped jointly by the sensory evidence received so far (exogenous drive) and the state propagated from earlier activity (endogenous dynamics), not by stimuli yet to arrive \cite{friston2005theory,hasson2008hierarchy,honey2012slow,murray2014hierarchy}.

Recent multimodal encoders such as TRIBE \cite{d2025tribe}, TRIBE v2 \cite{d2026foundation}, and MIRAGE \cite{gokce2026mirage} have substantially advanced naturalistic fMRI encoding by aligning large visual, auditory, and linguistic representations with whole-brain responses, but they remain primarily stimulus-to-response models with future-stimulus leakage into current prediction. Their objectives optimize the prediction of fMRI signals from sensory features, rather than the learning of a latent brain state whose transition is itself optimized for recursive simulation.

This distinction matters because high encoding accuracy alone does not define a brain world model. A simulator of movie-evoked brain dynamics must carry forward an endogenous state: one that summarizes prior neural activity, incorporates only temporally admissible sensory evidence, and remains usable when its own predictions are recursively fed back into the transition. The central question therefore shifts from which movie features best reconstruct fMRI responses to which latent variables form a transition-sufficient state space for causal brain-state evolution.

Concurrent work BrainVista \cite{yin2026brainvista} advances causal naturalistic fMRI modeling by forecasting future brain tokens under a Stimulus-to-Brain mask that prevents future-stimulus leakage. Its limitation lies not in temporal validity, but in two coupled design choices. First, its brain tokens are learned through fMRI reconstruction before forecasting. This is less ideal because for high-dimensional and noisy fMRI signals, this objective can spend latent capacity on current-sample detail, measurement fluctuations, and weakly predictable components, features that improve reconstruction but do not support future evolution \cite{dong2024brain}. Second, BrainVista interleaves stimulus and brain tokens within a shared autoregressive sequence and trains them through next-token prediction, rather than explicitly factoring the model into separate state representations and stimulus-conditioned dynamics. As a result, observation coding and dynamics learning remain entangled: the same representation must both preserve the measured fMRI signals and support autoregressive rollout. It is therefore not optimized specifically to yield transition-stable state evolution under sensory drive.

Beyond these methodological limitations, the empirical basis of naturalistic brain encoding remains narrow. Recent studies largely reuse a small set of public cohorts: Algonauts 2025 \cite{gifford2024algonauts} and CineBrain \cite{gao2025cinebrain}, the two continuous movie-fMRI benchmarks adopted here, provide rich within-subject recordings but contain only four and six participants, respectively. To substantially broaden this empirical basis, our group newly collected Singapore Multimodal Imaging \& Naturalistic Dataset (SG-MIND), a naturalistic audiovisual benchmark comprising 20 participants - five times the cohort size of Algonauts and more than three times that of CineBrain (Figure~\ref{fig2}). Each participant contributed approximately seven hours of fMRI data during naturalistic audiovisual viewing. Its unified acquisition and processing protocol, together with a common pool of 291 clips presented across participants, provides population breadth, stimulus diversity, and controlled cross-subject comparability, enabling a more stringent assessment of whether learned brain dynamics remain stable across individuals and naturalistic contexts.

To address these challenges, we propose \textbf{NeuroWorld}, a stimulus-conditioned world model that simulates human brain dynamics within a causal latent space. NeuroWorld explicitly disentangles endogenous brain-state evolution from exogenous sensory drive and optimizes its latent representation directly for recursive transitions rather than the reconstruction of fMRI. This formulation enables temporally causal and transition-stable rollouts of whole-brain activity under continuous naturalistic stimulation. Our contributions are fourfold:

\begin{itemize}
    \item \textbf{A latent-space world model of human brain dynamics.}
    To our knowledge, NeuroWorld is the first world model to cast naturalistic fMRI prediction as stimulus-conditioned state evolution in a learned latent space of brain dynamics. Its two-stage framework first learns a transition-sufficient latent state through causal next-state prediction and then decodes recursively rolled-out latent trajectories into subject-specific, whole-brain fMRI responses. Crucially, the latent dynamics are learned without an explicit objective to reconstruct individual fMRI observations, allowing the representation to be optimized specifically for stable recursive prediction.

    \item \textbf{Large-scale evaluation across subjects and datasets.}
    We evaluate NeuroWorld on three movie-fMRI benchmarks comprising 30 participants, constituting, to our knowledge, the broadest subject-level evaluation of causal naturalistic brain forecasting to date. Central to this evaluation is \textbf{SG-MIND}, our newly collected benchmark of 20 participants, 8,519 paired fMRI--stimulus clips, and 140.7 person-hours of audiovisual experience, which substantially expands the population breadth and cross-subject comparability of existing benchmarks.

    \item \textbf{State-of-the-art causal rollout performance.}
    Under a strictly causal evaluation protocol that precludes access to future stimuli, NeuroWorld achieves state-of-the-art multi-step rollout performance across all evaluated benchmarks. It consistently retains its advantage at both short and long forecasting horizons, demonstrating greater robustness to autoregressive error accumulation.

    \item \textbf{Comprehensive interpretability of the learned dynamics.}
    Beyond predictive performance, we conduct extensive interpretability analyses. Specifically, we characterize the structure of the learned latent state space and the organization of the resulting dynamics across functional brain systems. These analyses provide insight into how NeuroWorld represents and propagates brain activity during naturalistic stimulation.
\end{itemize}

\begin{figure*}[!t]
\centerline{\includegraphics[width=\textwidth]{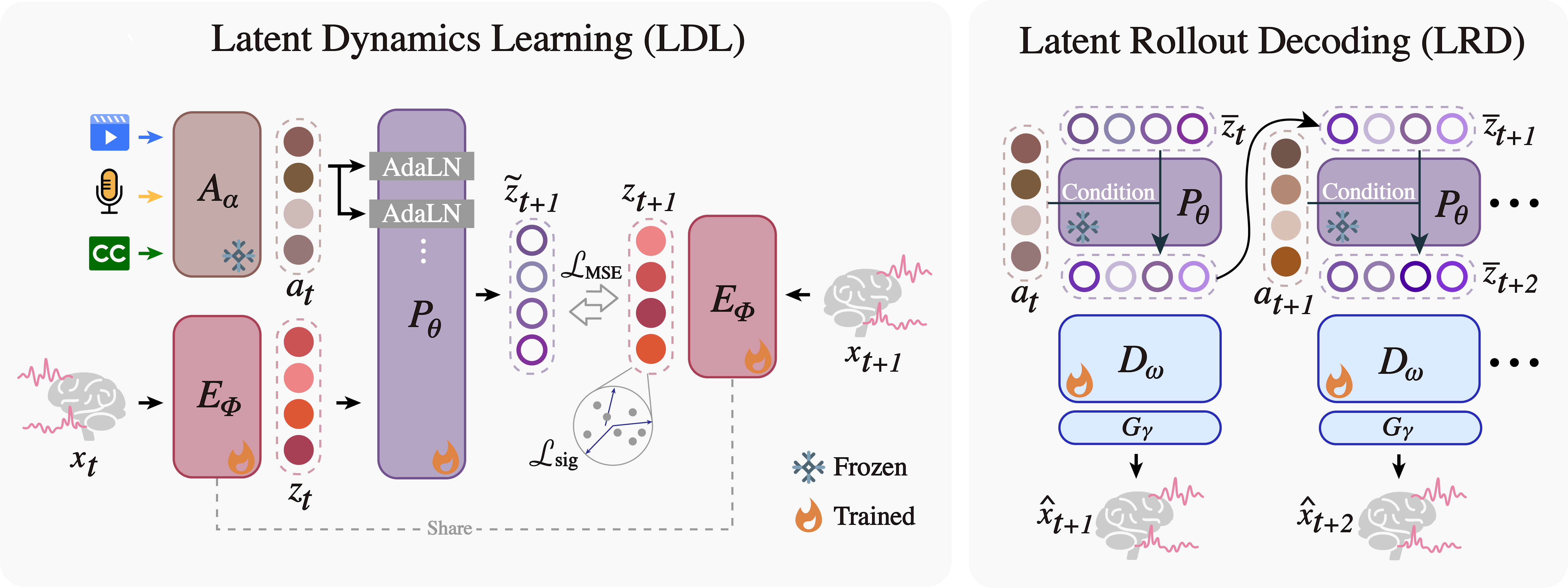}}
\caption{\textbf{Overview of NeuroWorld's two-stage framework.} \textbf{Left:} Latent Dynamics Learning (LDL) jointly learns the fMRI encoder $E_{\phi}$ and predictor $P_{\theta}$ through next-latent prediction with SIGReg, without an fMRI reconstruction objective. \textbf{Right:} Latent Rollout Decoding (LRD) freezes the learned world model, autoregressively rolls latent states forward under causal stimulus conditioning, and trains a shared decoder $D_{\omega}$ with subject-specific heads $G_{\gamma}$ to predict future whole-brain responses.}
\label{fig3}
\end{figure*}

\section{Related Work}
\label{sec:related_work}

\subsection{Naturalistic Brain Encoding Models}
\label{subsec:brain_encoding_models}

Naturalistic brain encoding aims to predict fMRI responses from temporally aligned sensory representations. TRIBE combines frozen video, audio, and language foundation models with a temporal Transformer shared across subjects and subject-conditioned fMRI readouts \cite{d2025tribe}; TRIBE v2 scales this formulation across heterogeneous datasets, increases the spatial resolution of its predictions, and supports zero-shot and fine-tuned prediction for unseen subjects \cite{d2026foundation}. MIRAGE instead extracts layer-resolved representations from a native multimodal backbone and learns adaptive feature gating before a shared temporal encoder and subject-specific readout \cite{gokce2026mirage}. Despite these advances, all three optimize direct stimulus-to-fMRI regression with future leakage, and their internal representations are selected for response fitting rather than for forming a dynamical state that remains valid under causal prediction. BrainVista moves closer to a world model by applying a Stimulus-to-Brain causal mask and autoregressively forecasting brain tokens \cite{yin2026brainvista}. Its fMRI tokenizer is nevertheless pretrained through observation reconstruction, while stimulus and brain tokens are interleaved within a single next-token stream; observation coding, sensory conditioning, and state evolution therefore remain coupled. NeuroWorld instead separates the endogenous brain state from the exogenous stimulus action and directly optimizes their causal latent transition before decoding the recursively rolled-out states.

\subsection{Foundation Models for Brain Dynamics}
\label{subsec:foundation_models_brain_dynamics}

Large-scale fMRI foundation models learn transferable representations of brain dynamics before downstream adaptation. BrainLM pretrains a masked autoencoder by reconstructing spatiotemporal fMRI patches \cite{ortega2024brainlm}. Brain-JEPA replaces signal reconstruction with joint-embedding prediction \cite{dong2024brain}. BrainHarmonix extends this formulation with cortical geometric harmonics and Temporal Adaptive Patch Embedding for heterogeneous repetition time (TR) \cite{dong2026brain}. Complementary approaches scale voxel-level 4D pretraining with a Mamba-based architecture, as in NeuroSTORM \cite{wang2025towards}. These models learn strong general-purpose representations, but their dynamics objectives condition only on neuroimaging observations and do not impose a past-only, stimulus-conditioned transition trained for recursive rollout. Concurrent work, BrainWorld, is more generative: it uses a reconstruction-trained VAE to obtain fMRI latents and a conditional Diffusion Transformer to predict and autoregressively roll out future frames from previous fMRI, structural MRI, functional connectivity, and optional video and audio conditions \cite{xia2026brainworld}. However, BrainWorld generates each target block jointly while allowing unrestricted cross-attention to the entire temporally aligned stimulus interval. It therefore does not enforce causality at the individual TR, and its recurrent state remains tied to a reconstructable observation latent. In contrast, our NeuroWorld learns a transition-sufficient state directly through next-latent prediction and incorporates sensory evidence only when that information becomes temporally available during rollout.

\section{Methodology}
\label{sec:method}

\subsection{Overview}
\label{subsec:method_overview}

Given a naturalistic movie and the corresponding fMRI, NeuroWorld learns a causal latent-space world model of brain dynamics. At each fMRI repetition time (TR), the ROI-level fMRI vector is encoded into a latent brain state, and the temporally corresponding multimodal stimulus features are encoded into a stimulus-action token. Stage~1, Latent Dynamics Learning (LDL), trains an action-conditioned latent transition model that predicts the next latent brain state from previous latent states and past-aligned stimulus actions. LDL is optimized by next-latent prediction, without any fMRI reconstruction loss. Stage~2, Latent Rollout Decoding (LRD), freezes the learned LDL, rolls the latent state forward autoregressively from an observed fMRI prefix, and trains a decoder that maps the rolled-out latent trajectory back to fMRI (Figure~\ref{fig3}).

\subsection{Problem Formulation}
\label{subsec:problem_formulation}

Let
$
    \mathbf{x}_{1:T}
    =
    \{\mathbf{x}_1,\ldots,\mathbf{x}_T\}
$
denote an fMRI sequence sampled on the TR grid, where $\mathbf{x}_t \in \Rbb^{R}$ is the ROI-level response at time $t$ and $R$ is the number of ROIs. Let $\Mset$ be the set of available stimulus modalities, such as video, audio, and text. The modality set is dataset-dependent, and the model only requires that each available modality can be represented as a time-indexed feature stream. The stimulus stream is treated as externally observed: the model conditions on stimulus actions but does not predict future stimuli.

The goal is to learn a causal transition in latent brain space. For each TR, an fMRI encoder $E_{\phi}$ maps the measured brain response to a latent state,
\begin{equation}
    \mathbf{z}_t = E_{\phi}(\mathbf{x}_t) \in \Rbb^{d_z},
    \label{eq:fmri_state_encoding}
\end{equation}
and an action encoder $A_{\alpha}$ maps the temporally aligned multimodal stimulus features to a stimulus-conditioned action token,
\begin{equation}
    \mathbf{a}_t
    =
    A_{\alpha}\!\left(\{\mathbf{s}_{m,t}\}_{m \in \Mset}\right)
    \in \Rbb^{d_a},
    \label{eq:stimulus_action_encoding}
\end{equation}
where $\mathbf{s}_{m,t}$ is the TR-aligned feature of modality $m$ at time $t$, $d_z$ is the latent brain-state dimension, and $d_a$ is the action-token dimension. A causal predictor $P_{\theta}$ parameterizes the latent transition using only temporally admissible information,
\begin{equation}
    \tilde{\mathbf{z}}_{t+1 \mid :t}
    =
    P_{\theta}
    \left(
    \mathbf{z}_{1:t},
    \mathbf{a}_{1:t}
    \right),
    \label{eq:causal_transition}
\end{equation}
where $\tilde{\mathbf{z}}_{t+1 \mid :t}$ denotes the predicted latent state for time $t+1$ conditioned on history up to time $t$. This formulation separates the endogenous state trajectory $\mathbf{z}_{1:t}$ from the exogenous stimulus drive $\mathbf{a}_{1:t}$. The final prediction target is the observable ROI-level fMRI response; the subsequent sections describe how the latent transition is learned in Stage~1 and how rolled-out latent states are decoded back to fMRI in Stage~2.

\subsection{Stage 1: Latent Dynamics Learning (LDL)}
\label{subsec:latent_dynamics_learning}

Latent Dynamics Learning (LDL) trains the latent state space and the causal transition jointly (Figure~\ref{fig3} left). Given a mini-batch of length-$L$ sequences, we first encode each measured fMRI TR and its aligned stimulus features as
\begin{equation}
    \mathbf{z}_{b,t}
    =
    E_{\phi}(\mathbf{x}_{b,t}),
    \qquad
    \mathbf{a}_{b,t}
    =
    A_{\alpha}\!\left(\{\mathbf{s}_{b,m,t}\}_{m\in\Mset}\right).
\end{equation}
where $b$ indexes the sequence in the mini-batch. The transition model then predicts the next latent state at every valid position by teacher forcing:
\begin{equation}
    \tilde{\mathbf{z}}_{b,t+1 \mid :t}
    =
    P_{\theta}
    \left(
    \mathbf{z}_{b,1:t},
    \mathbf{a}_{b,1:t}
    \right),
    \qquad
    1 \leq t < L .
    \label{eq:ldl_teacher_forcing}
\end{equation}
Within $P_{\theta}$, stimulus actions are injected through adaptive layer normalization (AdaLN), with $\mathbf{a}_{b,t}$ parameterizing feature-wise scale and shift terms that modulate the latent activations in each predictor block \cite{peebles2023scalable}. Here teacher forcing means that the history $\mathbf{z}_{b,1:t}$ is obtained from measured fMRI through $E_{\phi}$, rather than from the model's own previous predictions. The predictor is causal along the TR axis, so the output at position $t$ can attend only to latent states and stimulus-action tokens up to $t$. The loss is always applied to the final post-transition prediction $\tilde{\mathbf{z}}_{b,t+1 \mid :t}$.

The stimulus-action tokens are constructed on the fMRI TR grid with the
hemodynamic offset applied during sample construction. This
offset may require stimulus indices before the beginning of a run for windows
near the boundary; these positions are zero-padded and marked invalid. Let
$q_{b,t}\in\{0,1\}$ denote whether the action token at position $t$ in sequence
$b$ is a real shifted stimulus token rather than boundary padding. The valid
index set for LDL is
\begin{equation}
    \mathcal{I}_{\mathrm{LDL}}
    =
    \left\{
    (b,t): q_{b,t}=1,\ 1 \leq t < L
    \right\}.
\end{equation}
The mask is aligned with the prediction side, because the prediction at position
$t$ is conditioned on $\mathbf{a}_{b,t}$ and is trained to predict
$\mathbf{z}_{b,t+1}$. The next-latent prediction loss is
\begin{equation}
    \mathcal{L}_{\mathrm{pred}}(\phi,\theta)
    =
    \frac{1}{|\mathcal{I}_{\mathrm{LDL}}|}
    \sum_{(b,t)\in\mathcal{I}_{\mathrm{LDL}}}
    \left\|
    \tilde{\mathbf{z}}_{b,t+1 \mid :t}
    -
    \mathbf{z}_{b,t+1}
    \right\|_2^2 .
    \label{eq:ldl_prediction_loss}
\end{equation}
The target $\mathbf{z}_{b,t+1}$ is the latent encoding of the next measured fMRI TR. The fMRI encoder and the transition model are optimized jointly.

Jointly learning the encoder and transition can admit degenerate latent solutions. To prevent latent collapse, LDL regularizes the encoder outputs with Sketched Isotropic Gaussian Regularization (SIGReg) \cite{balestriero2025lejepa,maes2026leworldmodel}. Let $\{\mathbf{r}_j\}_{j=1}^{Q}$ be random unit projection vectors in $\Rbb^{d_z}$, and let $\mathcal{S}_{\mathrm{sig}}(\cdot)$ denote the one-dimensional normality statistic used by SIGReg. The regularizer is
\begin{equation}
    \mathcal{L}_{\mathrm{sig}}(\phi)
    =
    \frac{1}{Q}
    \sum_{j=1}^{Q}
    \mathcal{S}_{\mathrm{sig}}
    \left(
    \left\{
    \mathbf{r}_j^{\top}\mathbf{z}_{b,t}
    \right\}_{b,t}
    \right).
    \label{eq:ldl_sigreg}
\end{equation}
SIGReg is applied to encoder latents $\mathbf{z}_{b,t}$ because these latents define the recurrent state space used by the world model. The full LDL objective is
\begin{equation}
    \begin{aligned}
        (\phi^{\star},\theta^{\star})
        &=
        \arg\min_{\phi,\theta}
        \mathcal{L}_{\mathrm{LDL}}(\phi,\theta), \\
        \mathcal{L}_{\mathrm{LDL}}(\phi,\theta)
        &=
        \mathcal{L}_{\mathrm{pred}}(\phi,\theta)
        +
        \lambda
        \mathcal{L}_{\mathrm{sig}}(\phi).
    \end{aligned}
    \label{eq:ldl_objective}
\end{equation}
where $\lambda$ controls the strength of the latent regularizer. LDL contains no fMRI reconstruction term, so the latent representation is optimized for transition-stable future prediction rather than for reconstructing the current measured sample.

\subsection{Stage 2: Latent Rollout Decoding (LRD)}
\label{subsec:latent_rollout_decoding}

Latent Rollout Decoding (LRD) converts the learned latents into ROI-level fMRI predictions (Figure~\ref{fig3} right). After LDL, the encoder $E_{\phi}$ and transition model $P_{\theta}$ are frozen. LRD then trains only a decoder that maps rolled-out latent states back to fMRI space.

Let $H_0$ denote the number of observed fMRI TRs used to initialize the rollout, and let $H$ denote the rollout horizon. The initial latent prefix is obtained from measured fMRI,
\begin{equation}
    \bar{\mathbf{z}}_{b,t}
    =
    E_{\phi}(\mathbf{x}_{b,t}),
    \qquad
    1 \leq t \leq H_0 .
    \label{eq:lrd_prefix_encoding}
\end{equation}
where $\bar{\mathbf{z}}_{b,t}$ denotes the latent state used during rollout. Future latent states are generated autoregressively by feeding the model's own predictions back into the transition:
\begin{equation}
    \bar{\mathbf{z}}_{b,t+1}
    =
    P_{\theta}
    \left(
    \bar{\mathbf{z}}_{b,1:t},
    \mathbf{a}_{b,1:t}
    \right),
    \qquad
    H_0 \leq t < H_0 + H .
    \label{eq:lrd_autoregressive_rollout}
\end{equation}
The stimulus-action sequence remains externally observed throughout rollout. NeuroWorld therefore simulates only the latent brain trajectory, not the future stimulus stream.

For subject $s$, LRD uses a shared decoder trunk followed by a
subject-specific linear head. Let $D_{\omega}$ denote the shared decoder trunk,
and let $G_{\gamma_s}$ denote the linear head for subject $s$. For a sequence
$b$ from subject $s$, the fMRI prediction is
\begin{equation}
    \hat{\mathbf{x}}_{b,t}
    =
    G_{\gamma_s}
    \left(
    D_{\omega}
    \left(
    \bar{\mathbf{z}}_{b,t}
    \right)
    \right)
    \in \Rbb^{R}.
    \label{eq:lrd_decoder}
\end{equation}
The latent world model remains frozen and shared across
subjects; subject-specificity is introduced only in the final readout.

The decoder is trained on the rollout segment rather than on teacher-forced
encoder latents. Let
\[
    \mathcal{I}_{\mathrm{LRD}}
    =
    \left\{
    (b,t): q_{b,t}=1,\ H_0 \leq t < L
    \right\}
\]
be the valid set of rollout prediction indices within a length-$L$ training
window. In the summation below, each prediction $\hat{\mathbf{x}}_{b,t}$ is
computed using the linear head associated with the subject of sequence $b$.
The LRD objective is masked ROI-level mean-squared error,
\begin{equation}
    \mathcal{L}_{\mathrm{LRD}}(\omega,\gamma)
    =
    \frac{1}{|\mathcal{I}_{\mathrm{LRD}}|}
    \sum_{(b,t)\in\mathcal{I}_{\mathrm{LRD}}}
    \left\|
    \hat{\mathbf{x}}_{b,t+1}
    -
    \mathbf{x}_{b,t+1}
    \right\|_2^2,
    \label{eq:lrd_objective}
\end{equation}
where $\gamma=\{\gamma_s:s\in\mathcal{S}\}$ denotes the collection of
subject-specific linear-head parameters over the subject set $\mathcal{S}$.
LRD optimizes
\begin{equation}
    (\omega^{\star},\gamma^{\star})
    =
    \arg\min_{\omega,\gamma}
    \mathcal{L}_{\mathrm{LRD}}(\omega,\gamma).
    \label{eq:lrd_optimization}
\end{equation}
Only the shared decoder trunk $D_{\omega}$ and the subject-specific linear
heads $\{G_{\gamma_s}\}_{s\in\mathcal{S}}$ are updated in LRD. The LDL
parameters, including $E_{\phi}$ and $P_{\theta}$, remain
frozen.

\section{Experiments and Results}
\label{sec:experiments}

\subsection{Experimental Setup}

\begin{table}[t]
\centering
\caption{Detailed comparison of three benchmarks.}
\label{tab1}
\setlength{\tabcolsep}{5pt}
\begin{tabular}{lccc}
\toprule
 & Algonauts 2025 & CineBrain & SG-MIND \\
\midrule
Stimuli Modalities  & Video+Audio+Text & Video+Audio & Video+Audio \\
Subjects            & 4        & 6        & 20       \\
Total person-hours  & $\sim$260 h & $\sim$36 h & 140.7 h \\
Per subject         & $\sim$65 h  & $\sim$6 h  & ~$\sim$7 h   \\
TR                  & 1.49 s   & 0.80 s   & 0.719 s  \\
Spatial Resolution  & 2 mm   & 2 mm   & 2.5 mm  \\
\bottomrule
\end{tabular}
\end{table}

\begin{table*}[t]
\centering
\caption{Main results under the causal setting with a horizon of 20 steps on SG-MIND, Algonauts 2025, and CineBrain. Best values are marked in bold.}
\label{tab2}
\setlength{\tabcolsep}{4pt}
\footnotesize
\begin{tabular}{l ccc ccc ccc ccc}
\toprule
\multirow{3}{*}{Model} & \multicolumn{3}{c}{SG-MIND} & \multicolumn{6}{c}{Algonauts 2025} & \multicolumn{3}{c}{CineBrain} \\
\cmidrule(lr){2-4} \cmidrule(lr){5-10} \cmidrule(lr){11-13}
& \multicolumn{3}{c}{Random (0.8:0.2)} & \multicolumn{3}{c}{Random (0.8:0.2)} & \multicolumn{3}{c}{Season 6} & \multicolumn{3}{c}{Random (0.8:0.2)} \\
\cmidrule(lr){2-4} \cmidrule(lr){5-7} \cmidrule(lr){8-10} \cmidrule(lr){11-13}
& r & Pearson & Cls@Top10 & r & Pearson & Cls@Top10 & r & Pearson & Cls@Top10 & r & Pearson & Cls@Top10 \\
\midrule
Linear & 0.0486 & 0.0362 & 0.1210 & 0.2722 & 0.2275 & 0.6338 & 0.2659 & 0.2234 & 0.6286 & 0.2400 & 0.2169 & 0.4857 \\
TRIBE based & 0.0716 & 0.0900 & 0.0824 & 0.2789 & 0.2535 & 0.6015 & 0.2726 & 0.2492 & 0.5909 & 0.0983 & 0.0914 & 0.1878 \\
MIRAGE based & 0.0725 & 0.0813 & 0.1163 & 0.2771 & 0.2535 & 0.6426 & 0.2775 & 0.2544 & 0.6517 & 0.0806 & 0.0859 & 0.2044 \\
BrainVista & - & - & - & - & - & - & 0.2183$^{*}$ & 0.2090$^{*}$ & 0.7619$^{*}$ & - & - & - \\
\addlinespace[2pt]
NeuroWorld & \textbf{0.2190} & \textbf{0.2107} & \textbf{0.8044} & \textbf{0.2961} & \textbf{0.2759} & \textbf{0.8462} & \textbf{0.2934} & \textbf{0.2729} & \textbf{0.8488} & \textbf{0.2928} & \textbf{0.2817} & \textbf{0.7289} \\
\bottomrule
\end{tabular}
\par\vspace{3pt}
\parbox{\textwidth}{\footnotesize $^{*}$BrainVista~\cite{yin2026brainvista} provides no public code, so it could not be reproduced under our protocol; its results are quoted from the original paper and included only for approximate reference.}
\end{table*}

\begin{table}[t]
\centering
\caption{Results on Algonauts 2025 with different horizon H.}
\label{tab3}
\setlength{\tabcolsep}{5pt}
\resizebox{\columnwidth}{!}{%
\begin{tabular}{l cccccc}
\toprule
\multirow{3}{*}{Horizon} & \multicolumn{6}{c}{Algonauts 2025} \\
\cmidrule(lr){2-7}
& \multicolumn{3}{c}{Random (0.8:0.2)} & \multicolumn{3}{c}{Season 6} \\
\cmidrule(lr){2-4} \cmidrule(lr){5-7}
& r & Pearson & Cls@Top10 & r & Pearson & Cls@Top10 \\
\midrule
20 TR  & 0.2961 & 0.2759 & 0.8462 & 0.2934 & 0.2729 & 0.8488 \\
50 TR  & 0.2701 & 0.2470 & 0.6747 & 0.2701 & 0.2465 & 0.6863 \\
100 TR & 0.2660 & 0.2412 & 0.6488 & 0.2630 & 0.2385 & 0.6610 \\
\bottomrule
\end{tabular}%
}
\end{table}

\subsubsection{SG-MIND} 20 participants (22-40 years; 9 females) were recruited for the Investigating Individual Brain Activity in Response to Visual Stimuli (SG-MIND) study. One participant withdrew from the study after one session. All participants had normal hearing and normal or corrected-to-normal vision. Written informed consent was obtained, and the experimental protocol received approval from the National University of Singapore (NUS) ethics committee (NUS-IRB-2023-141). We selected over 500 approximately one-minute video clips from two popular American TV series – The Big Bang Theory and Friends – and home-made videos as the audiovisual stimuli.

All participants except three (SG-MIND01, SG-MIND12, and SG-MIND17) completed all 103 runs in approximately 22-26 sessions. Each session comprised 3-5 runs, with each run containing 5 video trials. In total, each participant contributed over 7 hours of video-watching fMRI data. 

During the stimulus presentation, the video resolution was adjusted to 1760 $\times$ 990 ratio to match the MRI’s in-bore LCD screen. Visual stimuli was presented on an LCD screen positioned at the head of the MRI scanner bed. Participants viewed the screen through a mirror attached to the RF coil, maintaining fixation on a white central cross over a grey background.  Participants were provided with custom-designed non-magnetic headphones with soft padding to minimise scanner noise, ensure comfort and clear audio delivery.

MRI data were collected in a 3T Siemens Prisma-fit scanner at the Centre for Translational MR Research in the National University of Singapore using a 32‐channel head coil. Anatomical images were collected with a T1-weighted Magnetisation Prepared Rapid Gradient Recalled Echo sequence (MPRAGE; 176 sagittal slices, resolution 1 mm isotropic, repetition time = 1950 ms, echo time = 2.98 ms) and were used for co-registering the fMRI data and projecting the data to a surface representation. fMRI data during movie viewing were collected with a multi-band EPI sequence (TR/TE = 719/30 ms, voxel size= 2.5$\times$2.5$\times$2.5 mm$^3$, FOV = 220$\times$220 mm$^2$, 60 interleaved axial slices, flip angle = 50°, bandwidth = 2840 Hz/pixel, multiband acceleration factor = 6; anterior-posterior phase encoding direction). A single volume of fMRI data with a reverse phase encoding direction (posterior-anterior) was further collected for spatial distortion correction.

Anatomical T1 data and fMRI data were preprocessed using the 
\href{https://github.com/ThomasYeoLab/CBIG}{CBIG pipeline}. Briefly, T1 data were subject to standard Free Surfer preprocessing; fMRI data underwent slice timing correction, motion correction, spatial distortion correction, alignment with T1, nuisance regression (motion parameters, white matter, CSF, and global signal), high-pass filtering (0.009 Hz) to remove slow drifts, and projection to the standard fsaverage6 surface template. Nuisance effects were mitigated via volume censoring and interpolation \cite{power2014methods}.

\subsubsection{Benchmarks} Experiments were conducted on three naturalistic fMRI benchmarks: Algonauts 2025 \cite{gifford2024algonauts}, CineBrain \cite{gao2025cinebrain}, and SG-MIND, whose details and differences are summarized in Table~\ref{tab1}. For each benchmark, we constructed a random train/validation split of 80\% and 20\% under a fixed seed, ensuring that validation stimuli were held out from training to prevent data leakage. For Algonauts 2025, we also adopted a stricter split that holds out all Friends season-6 runs for validation and trains on the remaining seasons.

\subsubsection{Stimulus Feature Extraction} Following TRIBE \cite{d2025tribe}, we extracted stimulus features with frozen pretrained encoders: Video-JEPA 2 gigantic \cite{assran2025v} for video (1408-d), Wav2Vec-Bert-2.0 \cite{chung2021w2v} for audio (1024-d), and Llama-3.2-3B \cite{grattafiori2024llama} for text (3072-d). The intermediate layers were aggregated into two depth-wise group means spanning relative depths $0.5$--$0.75$ and $0.75$--$1.0$ (the shallower layers were discarded, following TRIBE), and the resulting embeddings were resampled onto a common $2$~Hz grid. To align the stimuli with the fMRI response, the features were then mean-pooled from $2$~Hz onto each benchmark's TR grid and shifted by the hemodynamic lag ($\Delta=3$~TR $\approx4.47$~s for Algonauts 2025 and $\Delta=6$~TR $\approx4.31$~s for SG-MIND, while CineBrain were already aligned and uses $\Delta=0$). Note that all three provide video and audio modalities, while only Algonauts 2025 also includes text.

\subsubsection{fMRI Network Partition} To unify the architecture and training procedure across benchmarks, CineBrain and SG-MIND were reprocessed into the same Schaefer-2018 1000-parcel parcellation used by Algonauts 2025 \cite{schaefer2018local}. For CineBrain, two parcels carry no signal after this conversion and were removed, leaving 998 ROIs. We then partitioned the parcels into seven functional networks \cite{yeo2011organization} following BrainVista \cite{yin2026brainvista}, which defines the per-network fMRI encoder.

\subsubsection{Training and Evaluation} For Stage 1, we trained NeuroWorld for 50 epochs using AdamW with a batch size of 16; the learning rate was warmed up linearly to $2\times10^{-4}$ over the first 100 steps and then decayed with a cosine schedule. For Stage 2, we froze the components in stage 1 and trained the decoder for 200 epochs using AdamW with a batch size of 16; the learning rate was warmed up linearly to $10^{-3}$ over the first 500 steps and then decayed with a cosine schedule.

At test time, NeuroWorld remained frozen: the rollout was initialized from a single observed fMRI TR, and the latent state was propagated autoregressively under causal stimulus conditioning for a horizon of $H$ TRs ($H=20$ by default). We evaluated the results with three metrics per run and averaged over all validation runs pooled across all subjects: \textbf{r}, the global Pearson correlation over the flattened space--time trajectory; \textbf{Pearson}, the mean per-parcel temporal correlation, \emph{i.e.}, the standard Algonauts encoding score; and \textbf{Cls@Top10}, the top-10 identification accuracy of 20-TR trajectory segments.

\subsection{Main Results}
The ``Linear'' baseline adapts the official Algonauts 2025 baseline, a linear regression that maps stimulus features directly onto responses, to the extracted multimodal features of each benchmark. As a static, non-temporal regressor, it does not perform autoregressive rollout and is therefore reported outside the causal protocol. All other methods were evaluated under the same causal setting. For those that are not causal by design we modified the transformer blocks in encoders with causal masks. All results are the average score of all subjects.


\begin{figure}[!t]
\centerline{\includegraphics[width=0.9\columnwidth]{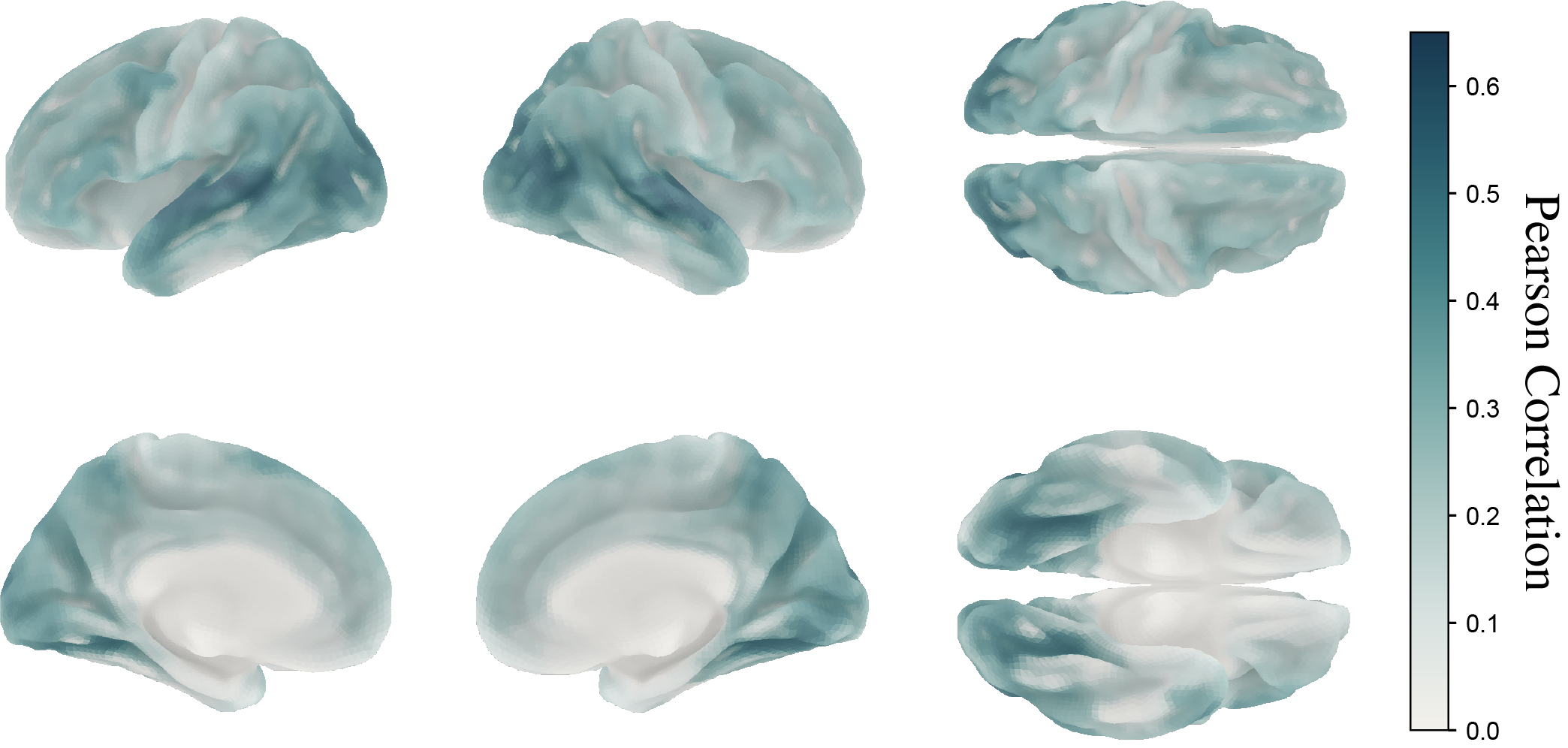}}
\caption{\textbf{Cortical distribution of NeuroWorld's prediction performance on Algonauts 2025}, with higher correlations concentrated in bilateral occipital and posterior temporal regions.}
\label{exp-1}
\end{figure}

Table~\ref{tab2} reports 20-step causal rollout performance under the three metrics defined above ($r$, Pearson, and Cls@Top10). NeuroWorld attains the best value on every metric of every benchmark and split. On Algonauts 2025, where the regression baselines are strongest owing to abundant per-subject recordings and full trimodal stimuli, NeuroWorld still raises the encoding score from 0.2535 to 0.2759 on the random split and from 0.2544 to 0.2729 on the season-6 split; the margin widens sharply on the identification metric, rising to 0.8462 from at most 0.6426 on the random split and to 0.8488 from at most 0.7619 on the season-6 split. The contrast is far more pronounced on the two harder benchmarks. On CineBrain, the causally masked TRIBE- and MIRAGE-based models collapse to $r\approx0.08$--$0.10$, falling below even the linear baseline ($r=0.2400$), whereas NeuroWorld remains stable at 0.2928. On SG-MIND, our largest and most diverse cohort, all regression baselines degrade toward chance ($r\le0.073$, Cls@Top10 $\le0.121$), while NeuroWorld retains $r=0.2190$ and Cls@Top10 $=0.8044$. Two observations stand out. First, the gap between NeuroWorld and the baselines is consistently largest on Cls@Top10, indicating that NeuroWorld produces trajectories that are not merely correlated on average but temporally and spatially discriminative. Second, once bidirectional stimulus access is removed and predictions are rolled out autoregressively, the strong stimulus-to-response fitting capacity of TRIBE and MIRAGE no longer translates into accurate forecasts; NeuroWorld's advantage grows precisely on the benchmarks with shorter per-subject data and fewer modalities (CineBrain, SG-MIND), where robust dynamics matter most, confirming that encoding accuracy alone does not yield a causally valid brain simulator.

Table~\ref{tab3} examines how NeuroWorld's rollout varies as the forecasting horizon $H$ increases from 20 to 100 TR on Algonauts 2025. As expected for autoregression, accumulated single-step error lowers every metric at longer horizons, but the decline is gradual for the correlation measures: on the random split $r$ falls only from 0.2961 to 0.2660 and the encoding score from 0.2759 to 0.2412, a relative drop of roughly 10--13\% over a fivefold horizon increase. The identification metric is more horizon-sensitive (Cls@Top10 from 0.8462 to 0.6488), because longer rollouts accumulate more drift before each segment is matched against its true counterpart. Crucially, the random and season-6 splits track each other closely at every horizon (e.g., $r=0.2660$ versus 0.2630 at $H=100$, and Cls@Top10 $=0.8462$ versus 0.8488 at $H=20$). This indicates that the learned dynamics generalize beyond the specific stimuli observed rather than overfitting to individual movie content, and that NeuroWorld's causal rollout remains stable over horizons well beyond the 20-TR training window.

\begin{figure}[!t]
\centerline{\includegraphics[width=0.9\columnwidth]{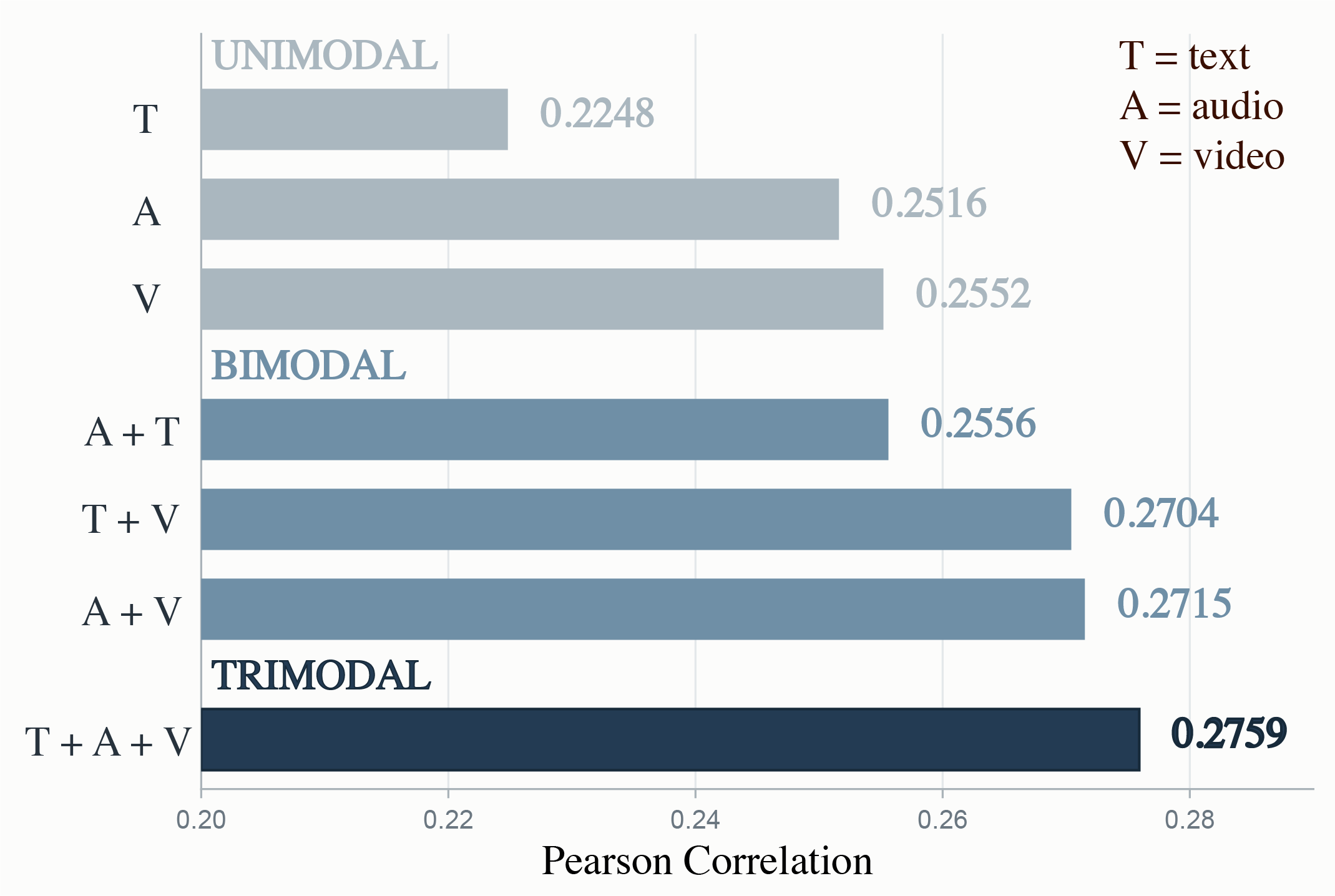}}
\caption{\textbf{Stimulus-modality ablation on Algonauts 2025.} Mean Pearson correlation under 20-step causal rollout for all combinations of text (T), audio (A), and video (V).}
\label{exp-2}
\end{figure}

\begin{figure}[!t]
\centerline{\includegraphics[width=0.95\columnwidth]{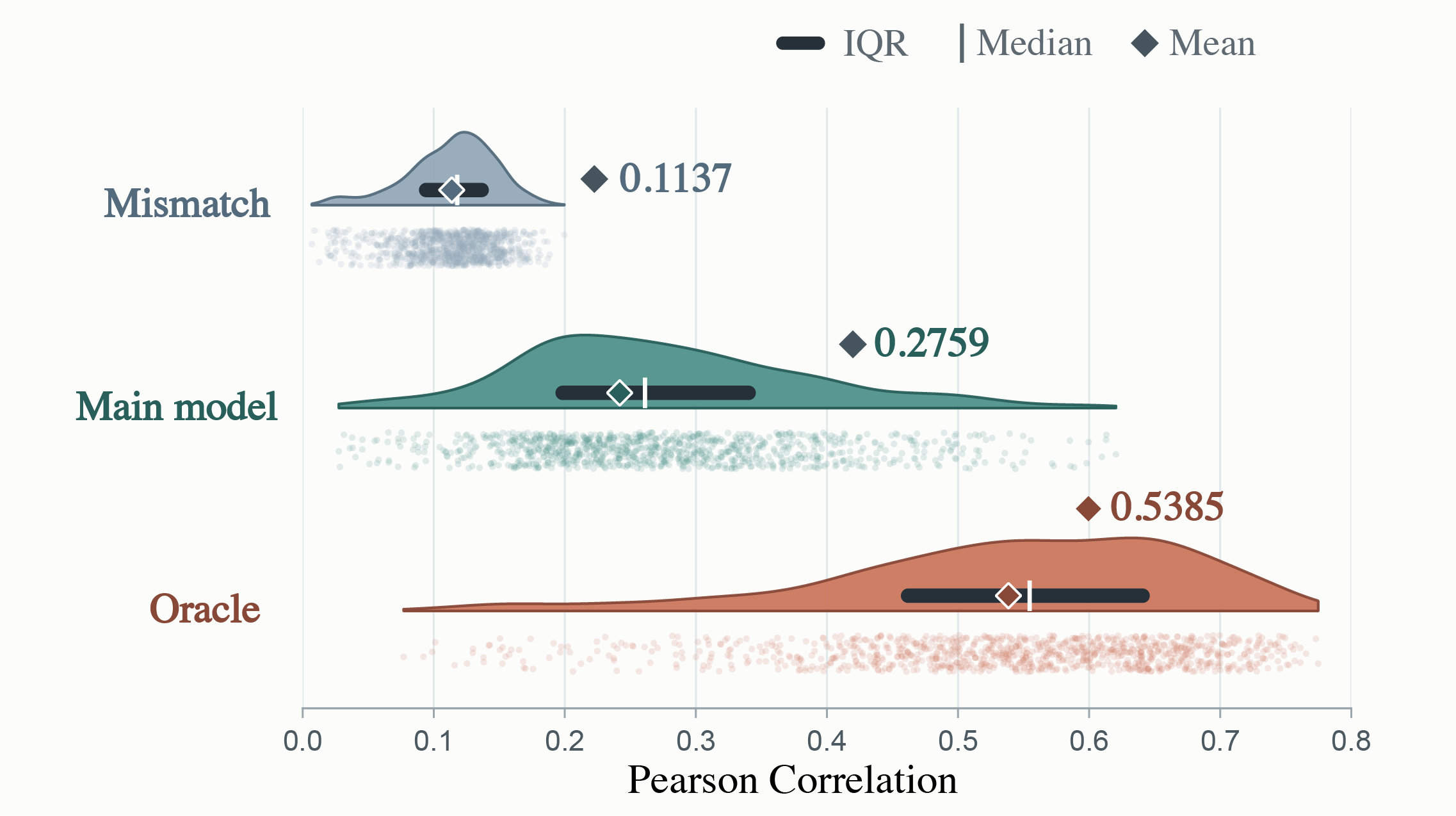}}
\caption{\textbf{Dissecting NeuroWorld's predictive power.} Aligned stimuli and accurate latent transitions jointly support autoregressive rollout. Pearson correlations for Mismatch, the main model (20-step autoregressive rollout), and the teacher-forced Oracle. Points denote ROIs.}
\label{exp-3}
\end{figure}

\begin{figure*}[!t]
\centerline{\includegraphics[width=\textwidth]{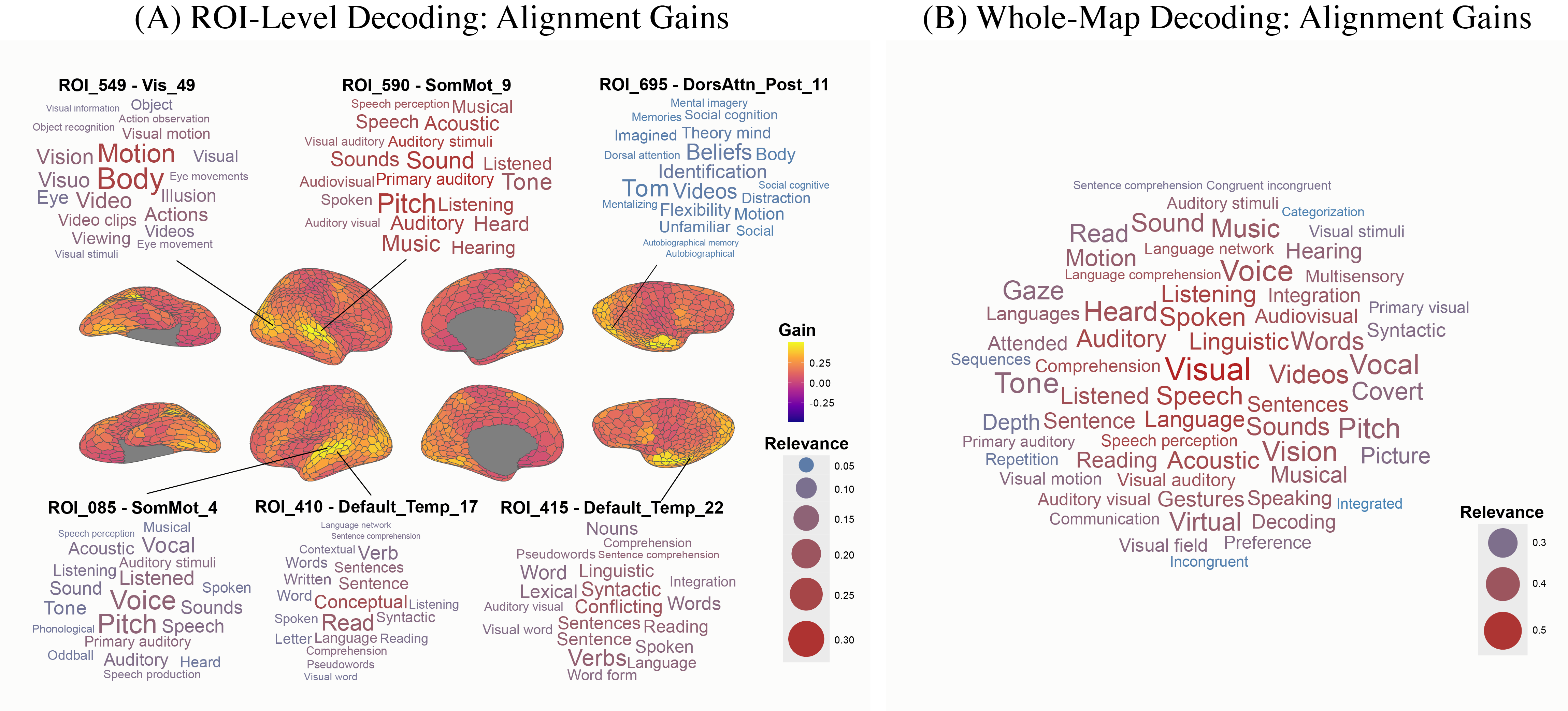}}
\caption{\textbf{Cognitive decoding of stimulus-brain alignment gains.} \textbf{A:} ROI-wise differences in Pearson's $r$ between aligned and Mismatch conditions, with the top 20 cognitive terms for representative high-gain parcels. \textbf{B:} Cognitive terms showing significant whole-map correspondence with the improvement topography (spin test, FDR $p<0.05$).}
\label{exp-4}
\end{figure*}

\subsection{Further Analysis}

\subsubsection{Cortical distribution of prediction performance}
Figure~\ref{exp-1} maps Pearson's $r$ across the Schaefer-1000 on Algonauts 2025. Prediction accuracy is bilateral and spatially widespread, with a clear posterior predominance. The strongest correlations are concentrated over medial and lateral occipital cortex and extend into posterior ventral occipitotemporal cortex, with additional elevations over posterior parietal and lateral temporal regions. This topography indicates particularly reliable prediction in visually responsive cortex, while the posterior temporal distribution is compatible with contributions from auditory and audiovisual processing regions \cite{beauchamp2004integration}. In contrast, correlations are comparatively weaker in anterior temporal, orbitofrontal, and medial prefrontal cortex. This pattern is consistent with the strong stimulus locking of visual and auditory cortices during naturalistic movie viewing and with the longer temporal integration windows of higher-order association cortex \cite{hasson2004intersubject}. Overall, NeuroWorld most reliably forecasts stimulus-coupled posterior cortical dynamics, whereas activity in anterior association regions remains more challenging to predict.

\subsubsection{Stimulus modality ablation}
To quantify the contribution of each stimulus stream, we vary the modalities provided to the stimulus-action encoder and evaluate all seven non-empty combinations of text (T), audio (A), and video (V) on Algonauts 2025. As shown in Figure~\ref{exp-2}, video achieves the highest unimodal Pearson correlation (0.2552), closely followed by audio (0.2516), whereas text alone yields lower performance (0.2248). Every bimodal combination outperforms its strongest constituent unimodal model, with audio--video performing best (0.2715), followed by text--video (0.2704) and audio--text (0.2556). Combining all three modalities further improves the correlation to 0.2759, corresponding to absolute gains of 0.0207 over the best unimodal variant and 0.0044 over the best bimodal variant. These results suggest that visual and auditory streams provide the primary predictive signals for movie-evoked brain dynamics, while text contributes complementary information beyond audiovisual conditioning.

\subsubsection{Dissecting NeuroWorld's predictive power}
We examine NeuroWorld's predictive power through three matched conditions (Figure~\ref{exp-3}). 1) Mismatch perserves the observed initial brain state but breaks stimulus--brain alignment; 2) the main model performs a 20-step autoregressive rollout with aligned multimodal actions; and 3) Oracle performs teacher-forced one-step prediction from the true latent history. Mean ROI-wise Pearson's $r$ increases from 0.1137 for Mismatch to 0.2759 for the main model and 0.5385 for Oracle. 

The more than twofold improvement over Mismatch shows that aligned sensory actions provide critical exogenous drive for steering the latent trajectory beyond endogenous state persistence and generic temporal structure. The strong Oracle performance further reveals high next-state predictive fidelity in the latent dynamics, while the main model transfers a substantial portion of this one-step capacity into sustained rollout from a single observed TR. Together, these results demonstrate NeuroWorld's ability to combine accurate local transitions with recursive simulation of extended brain dynamics.

\begin{figure*}[!t]
\centerline{\includegraphics[width=\textwidth]{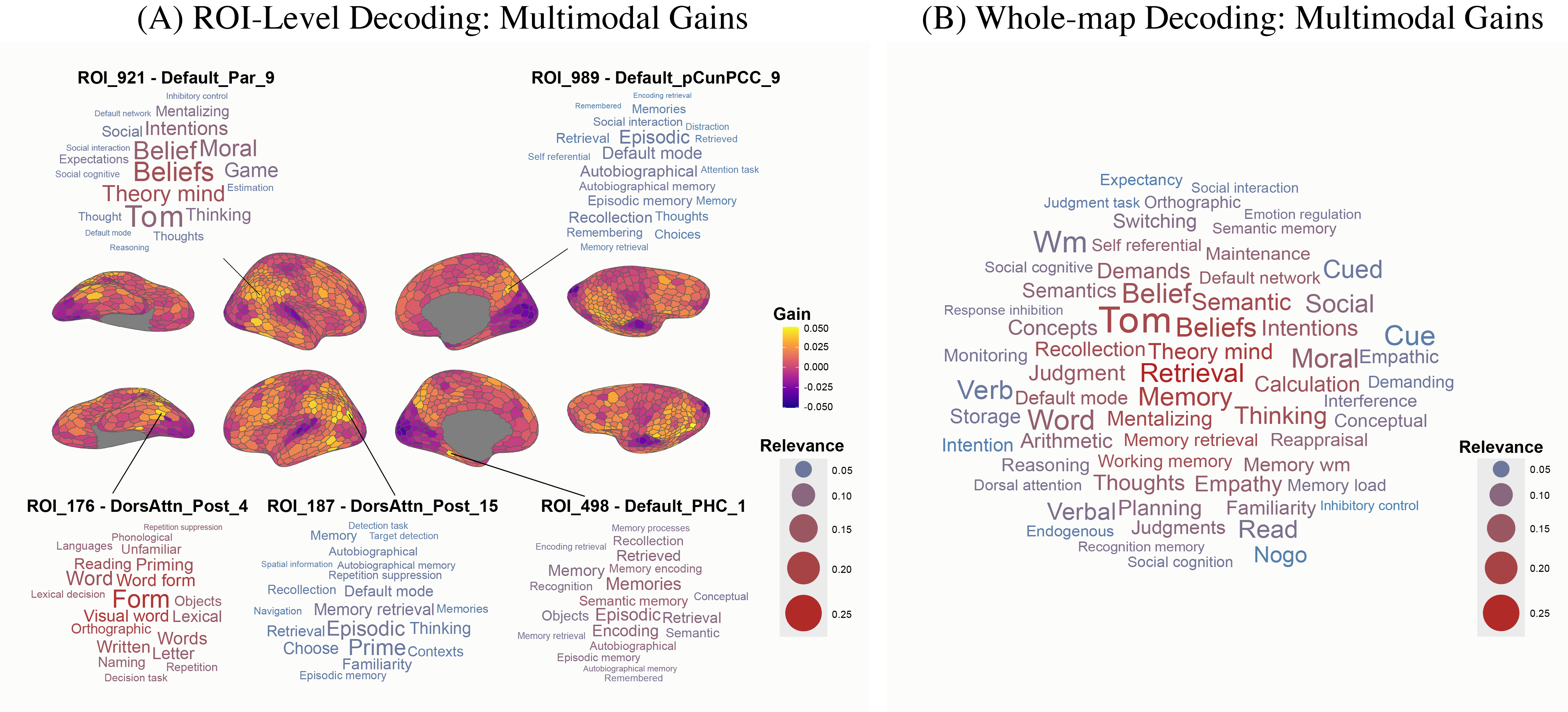}}
\caption{\textbf{Cognitive decoding of multimodal gains.} \textbf{A:} ROI-wise differences in Pearson's $r$ between the multimodal and best unimodal models, with the top 20 cognitive terms for representative high-gain parcels. \textbf{B:} Cognitive terms showing significant whole-map correspondence with the improvement topography (spin test, FDR $p<0.05$).}
\label{exp-5}
\end{figure*}

\begin{figure}[!t]
\centerline{\includegraphics[width=0.95\columnwidth]{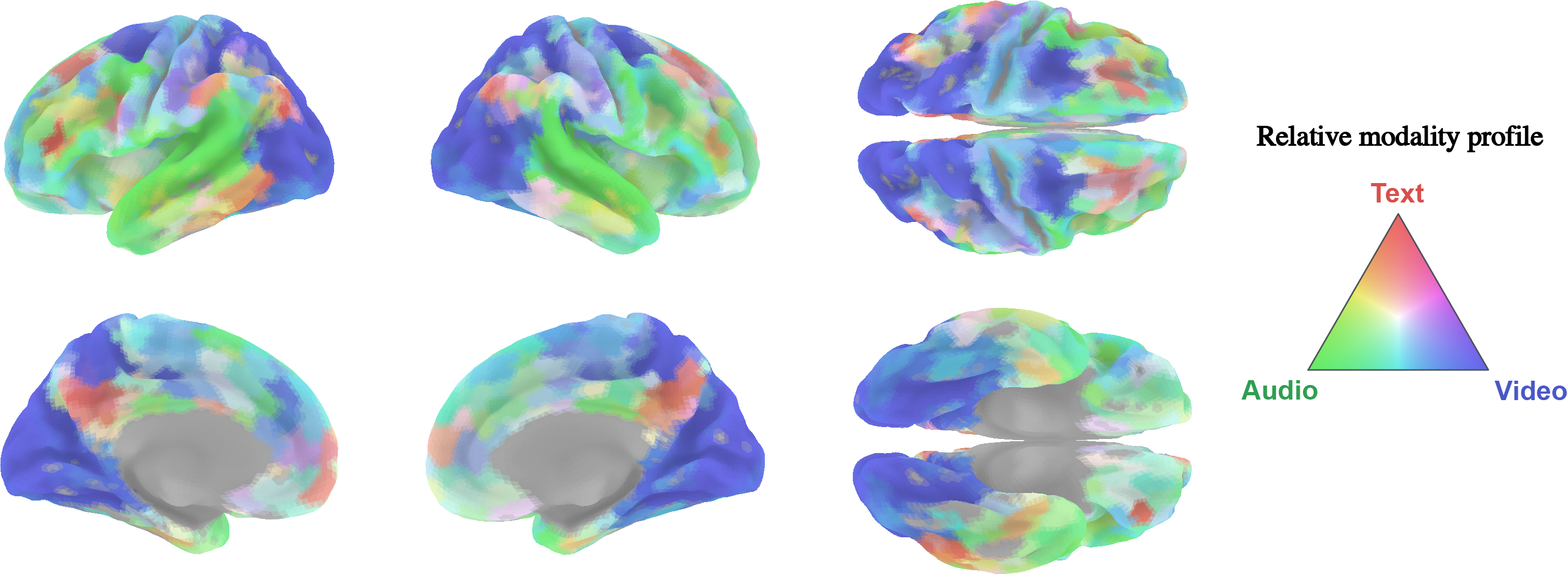}}
\caption{\textbf{Cortical mosaic of relative unimodal encoding profiles.} Colors encode relative ROI-wise predictive performance based on validation-set Pearson’s (r), for the text-only (red), audio-only (green), and video-only (blue) models. Blended colors indicate overlapping modality-specific predictive profiles.}
\label{exp-6}
\end{figure}

\subsubsection{Cognitive signatures of predictive gains}
To characterize the cognitive organization of NeuroWorld's predictive gains, we functionally decode two set of whole-brain ROI-wise delta-correlation $\Delta r$ maps (Pearson correlation differences) on Algonauts 2025: 1) correctly aligned versus mismatched stimulus actions (Figure \ref{exp-4}), and 2) multimodal versus best unimodal conditioning (Figure \ref{exp-5}). Functional decoding uses 506 Neurosynth-derived meta-analytic activation maps \cite{pacella2024morphospace}, cosine-similarity spin tests with 1,000 permutations and FDR correction ($p<0.05$) \cite{markello2022neuromaps}, and NiMARE-based ROI profiles \cite{salo2023nimare}. In each figure, the panel A presents the cortical $\Delta r$ map and the top 20 cognitive terms for representative high-gain ROIs, whereas the panel B summarizes terms showing significant whole-map correspondence. 

Alignment gains are concentrated in superior and middle temporal, inferior parietal, and occipital cortex, with the largest mean gains in the visual, dorsal attention, and default-mode networks. Both decoding levels emphasize visual motion, audition, speech, language, syntax, multisensory integration, and comprehension (Figure~\ref{exp-4}). 

Multimodal gains span prefrontal, temporal, and parietal association cortex, led by the default-mode, dorsal attention, and control networks, with decoded terms converging on episodic memory, mentalizing, semantic processing, and internally oriented cognition (Figure~\ref{exp-5}). Together, these contrasts indicate that stimulus--brain alignment grounds the rollout in the correct external event, whereas multimodal stimuli enriches the higher-order cognitive content of the latent brain state.

\subsubsection{Cortical mosaic of unimodal encoding}
Figure~\ref{exp-6} resolves the global modality ablation into a cortical map of relative unimodal encoding profiles. Each ROI blends the Pearson's $r$ of the text-, audio-, and video-only models according to their relative strengths. Video yields the highest unimodal correlation in 527 of 1,000 ROIs, including 160 of 162 visual-network parcels and 104 of 122 dorsal-attention parcels, forming a bilateral occipital-to-parietal profile. Audio is strongest in 369 ROIs, centered on superior temporal and perisylvian auditory cortex, consistent with distributed speech encoding across human auditory cortex \cite{hamilton2021parallel}. Broader preferences extend into somatomotor and salience/ventral-attention regions. Text is strongest in 104 ROIs, 85 within the default-mode and control networks. Its posterior-medial and frontoparietal distribution aligns with higher-order systems supporting semantic cognition, prediction, and integration during language comprehension \cite{fernandino2024does,he2025diverse}. Modality separation is most pronounced in the visual network and substantially weaker in the default-mode and control networks. This cortical mosaic reveals a progression from modality-specific sensory forecasting to increasingly shared predictive profiles in association cortex, providing a spatial account of the gains from trimodal conditioning.

\section{Conclusion and Discussion}
We introduced NeuroWorld, a stimulus-conditioned world model that reframes naturalistic fMRI encoding as causal evolution in a learned latent brain-state space. Its two-stage architecture separates latent dynamics learning from observation decoding: The Latent Dynamics Learning stage learns transition-sufficient brain states through next-latent prediction without reconstructing the observed fMRI signals. The Latent Rollout Decoding stage then maps recursively generated latent trajectories to subject-specific, whole-brain fMRI responses. 

Across Algonauts 2025, CineBrain, and the newly collected SG-MIND benchmark, comprising 30 participants in total, NeuroWorld achieves state-of-the-art causal rollout performance across all evaluated benchmarks. It further preserves substantial predictive accuracy over horizons of up to 100 TRs and generalizes to an entirely held-out movie season. Together, these findings support a central premise of this work: high retrospective encoding accuracy alone is insufficient to produce a brain simulator. Reliable forecasting requires state representations and transition objectives explicitly designed for temporal causality and recursive prediction, and stable rollout.

Our analyses further clarify the sources and functional organization of this predictive capacity. Correct stimulus-brain alignment yields a substantial improvement over mismatched conditioning, demonstrating that multimodal stimulus actions actively steer latent trajectories beyond the information retained in the initial brain state. At the same time, strong teacher-forced performance reveals high local predictive fidelity in the learned transition, a substantial portion of which is preserved during extended autoregressive rollout. Multimodal conditioning consistently outperforms unimodal alternatives: visual and auditory streams provide the dominant predictive signals, while text contributes complementary information in higher-order association systems. Spatial and meta-analytic analyses further reveal modality-selective predictive profiles in sensory cortices, increasingly shared multimodal representations across association networks, and distinct cognitive signatures associated with stimulus alignment and multimodal integration. NeuroWorld therefore provides not only an accurate world model of brain dynamics, but also a framework for investigating how endogenous brain states and exogenous sensory inputs jointly shape naturalistic cortical dynamics.

The present study only focuses on movie stimuli, ROI-level fMRI signals, held-out runs from previously observed cohorts, and subject-specific output heads. Important directions for future work include evaluation in independent cohorts and unseen participants, more efficient subject adaptation, longer and uncertainty-aware rollout, and extension to additional naturalistic paradigms and neuroimaging modalities. More broadly, controllable latent brain dynamics could support counterfactual analyses of how alternative sensory histories shape future neural trajectories. In conclusion, NeuroWorld represents a concrete step beyond static stimulus-to-response encoding toward causal, stateful simulation of the human brain during continuous naturalistic experience.

\section*{Acknowledgment}

This study was supported by the Singapore National Medical Research Council
(NMRC HLCA23Feb-0004, OFIRG24Jul-0049, Human Potential Joint Grant HPJGC25-0002), Human Potential Joint Grant H25P3M00001, HHP Industry-Alignment Fund H24J4a0143), Ministry of Education (MOE-T2EP20223-0013), and Yong Loo Lin School of Medicine Research Core Funding, National University of Singapore, Singapore.

\section*{References}
\bibliographystyle{IEEEtran}
\bibliography{refs}

\begin{thebibliography}{10}
\providecommand{\url}[1]{#1}
\csname url@samestyle\endcsname
\providecommand{\newblock}{\relax}
\providecommand{\bibinfo}[2]{#2}
\providecommand{\BIBentrySTDinterwordspacing}{\spaceskip=0pt\relax}
\providecommand{\BIBentryALTinterwordstretchfactor}{4}
\providecommand{\BIBentryALTinterwordspacing}{\spaceskip=\fontdimen2\font plus
\BIBentryALTinterwordstretchfactor\fontdimen3\font minus \fontdimen4\font\relax}
\providecommand{\BIBforeignlanguage}[2]{{%
\expandafter\ifx\csname l@#1\endcsname\relax
\typeout{** WARNING: IEEEtran.bst: No hyphenation pattern has been}%
\typeout{** loaded for the language `#1'. Using the pattern for}%
\typeout{** the default language instead.}%
\else
\language=\csname l@#1\endcsname
\fi
#2}}
\providecommand{\BIBdecl}{\relax}
\BIBdecl

\bibitem{d2025tribe}
S.~d'Ascoli, J.~Rapin, Y.~Benchetrit, H.~Banville, and J.-R. King, ``Tribe: Trimodal brain encoder for whole-brain fmri response prediction,'' \emph{arXiv preprint arXiv:2507.22229}, 2025.

\bibitem{d2026foundation}
S.~d'Ascoli, J.~Rapin, Y.~Benchetrit, T.~Brooks, K.~Begany, J.~Raugel, H.~Banville, and J.-R. King, ``A foundation model of vision, audition, and language for in-silico neuroscience,'' \emph{arXiv preprint arXiv:2605.04326}, 2026.

\bibitem{gokce2026mirage}
A.~Gokce, B.~AlKhamissi, and M.~Schrimpf, ``Mirage: Adaptive multimodal gating for whole-brain fmri encoding,'' \emph{arXiv preprint arXiv:2605.29850}, 2026.

\bibitem{friston2005theory}
K.~Friston, ``A theory of cortical responses,'' \emph{Philosophical transactions of the Royal Society B: Biological sciences}, vol. 360, no. 1456, p. 815, 2005.

\bibitem{hasson2008hierarchy}
U.~Hasson, E.~Yang, I.~Vallines, D.~J. Heeger, and N.~Rubin, ``A hierarchy of temporal receptive windows in human cortex,'' \emph{Journal of neuroscience}, vol.~28, no.~10, pp. 2539--2550, 2008.

\bibitem{honey2012slow}
C.~J. Honey, T.~Thesen, T.~H. Donner, L.~J. Silbert, C.~E. Carlson, O.~Devinsky, W.~K. Doyle, N.~Rubin, D.~J. Heeger, and U.~Hasson, ``Slow cortical dynamics and the accumulation of information over long timescales,'' \emph{Neuron}, vol.~76, no.~2, pp. 423--434, 2012.

\bibitem{murray2014hierarchy}
J.~D. Murray, A.~Bernacchia, D.~J. Freedman, R.~Romo, J.~D. Wallis, X.~Cai, C.~Padoa-Schioppa, T.~Pasternak, H.~Seo, D.~Lee \emph{et~al.}, ``A hierarchy of intrinsic timescales across primate cortex,'' \emph{Nature neuroscience}, vol.~17, no.~12, pp. 1661--1663, 2014.

\bibitem{yin2026brainvista}
X.~Yin, R.~Zhao, L.~Yao, and W.~Cai, ``Brainvista: Modeling naturalistic brain dynamics as multimodal next-token prediction,'' \emph{arXiv preprint arXiv:2602.04512}, 2026.

\bibitem{dong2024brain}
Z.~Dong, R.~Li, Y.~Wu, T.~T. Nguyen, J.~S. Chong, F.~Ji, N.~R. Tong, C.~L. Chen, and J.~H. Zhou, ``Brain-jepa: Brain dynamics foundation model with gradient positioning and spatiotemporal masking,'' \emph{Advances in Neural Information Processing Systems}, vol.~37, pp. 86\,048--86\,073, 2024.

\bibitem{gifford2024algonauts}
A.~T. Gifford, D.~Bersch, M.~St-Laurent, B.~Pinsard, J.~Boyle, L.~Bellec, A.~Oliva, G.~Roig, and R.~M. Cichy, ``The algonauts project 2025 challenge: How the human brain makes sense of multimodal movies,'' \emph{arXiv preprint arXiv:2501.00504}, 2024.

\bibitem{gao2025cinebrain}
J.~Gao, Y.~Liu, B.~Yang, J.~Feng, and Y.~Fu, ``Cinebrain: A large-scale multi-modal brain dataset during naturalistic audiovisual narrative processing,'' \emph{arXiv preprint arXiv:2503.06940}, 2025.

\bibitem{ortega2024brainlm}
J.~Ortega~Caro, A.~H. de~Oliveira~Fonseca, S.~Rizvi, M.~Rosati, C.~Averill, J.~Cross, P.~Mittal, E.~Zappala, R.~Dhodapkar, C.~Abdallah \emph{et~al.}, ``Brainlm: A foundation model for brain activity recordings,'' in \emph{International Conference on Learning Representations}, vol. 2024, 2024, pp. 565--576.

\bibitem{dong2026brain}
Z.~Dong, R.~Li, J.~Chong, N.~Dehestani, Y.~Teng, Y.~Lin, Z.~Li, Y.~Zhang, Y.~Xie, L.~Ooi \emph{et~al.}, ``Brain harmony: a multimodal foundation model unifying morphology and function into 1d tokens,'' \emph{Advances in Neural Information Processing Systems}, vol.~38, pp. 122\,100--122\,127, 2026.

\bibitem{wang2025towards}
C.~Wang, Y.~Jiang, Z.~Peng, C.~Li, C.~Bang, L.~Zhao, W.~Fu, J.~Lv, J.~Sepulcre, C.~Yang \emph{et~al.}, ``Towards a general-purpose foundation model for fmri analysis,'' \emph{arXiv preprint arXiv:2506.11167}, 2025.

\bibitem{xia2026brainworld}
J.~Xia, W.~Ye, J.~Zhang, X.~Pan, M.~Wang, and Q.~Liu, ``Brainworld: A structural-prior-conditioned generative model for whole-brain 4d fmri dynamics,'' \emph{arXiv preprint arXiv:2606.17742}, 2026.

\bibitem{peebles2023scalable}
W.~Peebles and S.~Xie, ``Scalable diffusion models with transformers,'' in \emph{Proceedings of the IEEE/CVF international conference on computer vision}, 2023, pp. 4195--4205.

\bibitem{balestriero2025lejepa}
R.~Balestriero and Y.~LeCun, ``Lejepa: Provable and scalable self-supervised learning without the heuristics,'' \emph{arXiv preprint arXiv:2511.08544}, 2025.

\bibitem{maes2026leworldmodel}
L.~Maes, Q.~L. Lidec, D.~Scieur, Y.~LeCun, and R.~Balestriero, ``Leworldmodel: Stable end-to-end joint-embedding predictive architecture from pixels,'' \emph{arXiv preprint arXiv:2603.19312}, 2026.

\bibitem{power2014methods}
J.~D. Power, A.~Mitra, T.~O. Laumann, A.~Z. Snyder, B.~L. Schlaggar, and S.~E. Petersen, ``Methods to detect, characterize, and remove motion artifact in resting state fmri,'' \emph{neuroimage}, vol.~84, pp. 320--341, 2014.

\bibitem{assran2025v}
M.~Assran, A.~Bardes, D.~Fan, Q.~Garrido, R.~Howes, M.~Muckley, A.~Rizvi, C.~Roberts, K.~Sinha, A.~Zholus \emph{et~al.}, ``V-jepa 2: Self-supervised video models enable understanding, prediction and planning,'' \emph{arXiv preprint arXiv:2506.09985}, 2025.

\bibitem{chung2021w2v}
Y.-A. Chung, Y.~Zhang, W.~Han, C.-C. Chiu, J.~Qin, R.~Pang, and Y.~Wu, ``W2v-bert: Combining contrastive learning and masked language modeling for self-supervised speech pre-training,'' in \emph{2021 IEEE Automatic Speech Recognition and Understanding Workshop (ASRU)}.\hskip 1em plus 0.5em minus 0.4em\relax IEEE, 2021, pp. 244--250.

\bibitem{grattafiori2024llama}
A.~Grattafiori, A.~Dubey, A.~Jauhri, A.~Pandey, A.~Kadian, A.~Al-Dahle, A.~Letman, A.~Mathur, A.~Schelten, A.~Vaughan \emph{et~al.}, ``The llama 3 herd of models,'' \emph{arXiv preprint arXiv:2407.21783}, 2024.

\bibitem{schaefer2018local}
A.~Schaefer, R.~Kong, E.~M. Gordon, T.~O. Laumann, X.-N. Zuo, A.~J. Holmes, S.~B. Eickhoff, and B.~T. Yeo, ``Local-global parcellation of the human cerebral cortex from intrinsic functional connectivity mri,'' \emph{Cerebral cortex}, vol.~28, no.~9, pp. 3095--3114, 2018.

\bibitem{yeo2011organization}
B.~T. Yeo, F.~M. Krienen, J.~Sepulcre, M.~R. Sabuncu, D.~Lashkari, M.~Hollinshead, J.~L. Roffman, J.~W. Smoller, L.~Z{\"o}llei, J.~R. Polimeni \emph{et~al.}, ``The organization of the human cerebral cortex estimated by intrinsic functional connectivity,'' \emph{Journal of neurophysiology}, 2011.

\bibitem{beauchamp2004integration}
M.~S. Beauchamp, K.~E. Lee, B.~D. Argall, and A.~Martin, ``Integration of auditory and visual information about objects in superior temporal sulcus,'' \emph{Neuron}, vol.~41, no.~5, pp. 809--823, 2004.

\bibitem{hasson2004intersubject}
U.~Hasson, Y.~Nir, I.~Levy, G.~Fuhrmann, and R.~Malach, ``Intersubject synchronization of cortical activity during natural vision,'' \emph{science}, vol. 303, no. 5664, pp. 1634--1640, 2004.

\bibitem{pacella2024morphospace}
V.~Pacella, V.~Nozais, L.~Talozzi, M.~Abdallah, D.~Wassermann, S.~J. Forkel, and M.~Thiebaut~de Schotten, ``The morphospace of the brain-cognition organisation,'' \emph{Nature Communications}, vol.~15, no.~1, p. 8452, 2024.

\bibitem{markello2022neuromaps}
R.~D. Markello, J.~Y. Hansen, Z.-Q. Liu, V.~Bazinet, G.~Shafiei, L.~E. Su{\'a}rez, N.~Blostein, J.~Seidlitz, S.~Baillet, T.~D. Satterthwaite \emph{et~al.}, ``Neuromaps: structural and functional interpretation of brain maps,'' \emph{Nature methods}, vol.~19, no.~11, pp. 1472--1479, 2022.

\bibitem{salo2023nimare}
T.~Salo, T.~Yarkoni, T.~E. Nichols, J.-B. Poline, M.~Bilgel, K.~L. Bottenhorn, D.~Jarecka, J.~D. Kent, A.~Kimbler, D.~M. Nielson \emph{et~al.}, ``Nimare: neuroimaging meta-analysis research environment,'' \emph{Aperture Neuro}, vol.~3, pp. 1--32, 2023.

\bibitem{hamilton2021parallel}
L.~S. Hamilton, Y.~Oganian, J.~Hall, and E.~F. Chang, ``Parallel and distributed encoding of speech across human auditory cortex,'' \emph{Cell}, vol. 184, no.~18, pp. 4626--4639, 2021.

\bibitem{fernandino2024does}
L.~Fernandino and J.~R. Binder, ``How does the “default mode” network contribute to semantic cognition?'' \emph{Brain and language}, vol. 252, p. 105405, 2024.

\bibitem{he2025diverse}
Y.~He, X.~Shao, C.~Liu, C.~Fan, E.~Jefferies, M.~Zhang, and X.~Li, ``Diverse frontoparietal connectivity supports semantic prediction and integration in sentence comprehension,'' \emph{Journal of neuroscience}, vol.~45, no.~5, 2025.

\end{thebibliography}

\end{document}